\documentclass[a4paper]{article}
\usepackage{listings}
\usepackage{xcolor}
\usepackage{multirow}
\usepackage{jheppub}  
\usepackage{dcolumn}
\usepackage[english]{babel}
\usepackage[utf8x]{inputenc}
\usepackage[T1]{fontenc}
\usepackage{palatino}
\usepackage{amsmath}
\usepackage{afterpage}
\usepackage{graphicx, subcaption}
\usepackage[colorinlistoftodos]{todonotes}
\usepackage[colorlinks=true]{hyperref}
\usepackage{makeidx}
\newcommand{\be}{\begin{equation}}  
\newcommand{\ee}{\end{equation}}  
\newcommand{\bea}{\begin{eqnarray}}  
\newcommand{\eea}{\end{eqnarray}}  
\makeindex
\begin{document}

\vspace*{1.2cm}

\thispagestyle{empty}
\begin{center}
{\LARGE \bf Exotic states with charm and/or strangeness}

\par\vspace*{7mm}\par

{

\bigskip

\large \bf A. Mart\'inez Torres${}^{1}$, Breno Agat\~ao${}^{1}$, Pedro Brand\~ao${}^{2}$, K. P. Khemchandani${}^{3}$, Luciano M. Abreu${}^{2}$, E. Oset${}^{4,5}$}

\bigskip

{\large \bf  E-Mail: amartine@if.usp.br}

\bigskip

{
${}^1$ Instituto de F\'{\i}sica, Universidade de S\~{a}o Paulo, Rua do Mat\~{a}o, São Paulo SP, 05508-090, Brazil.\\

${}^2$Instituto de Física, Universidade Federal da Bahia, Campus Ondina, Salvador, Bahia 40170-115, Brazil.\\

${}^3$ Universidade Federal de S\~ao Paulo, C.P. 01302-907, S\~ao Paulo, Brazil.\\
${}^4$ Departamento de F\'isica Te\'orica and IFIC, Centro Mixto Universidad de Valencia-CSIC, Institutos de Investigaci\'on de Paterna, 46071 Valencia, Spain.\\
${}^5$ Guangxi Key Laboratory of Nuclear Physics and Technology, Guangxi Normal University, Guilin 541004, China.}

\bigskip

{\it Presented at the Workshop of Advances in QCD at the LHC and the EIC, CBPF, Rio de Janeiro, Brazil, November 9-15 2025}


\vspace*{15mm}

\end{center}
\vspace*{1mm}

\begin{abstract}
~In this talk, I will discuss the properties of several exotic states with charm and/or strangeness. These states can be interpreted either as being generated from two- or three-hadron dynamics. In particular, I will focus on $D_1(2420)$ and some predictions of three-body states obtained from $n\bar D_{s1}(2460/2536)$. I will show how information on correlation functions and invariant mass distributions can reveal the exotic nature of these states.
\end{abstract}

 
 \section{Introduction}
 In recent years, the combined analysis of correlation functions and invariant mass distributions has become a powerful tool for investigating the properties of states that are not well described within the traditional quark model~\cite{ALICE:2022uso,Fabbietti:2020bfg,Khemchandani:2023xup,Albaladejo:2024lam,Abreu:2024qqo,Brandao:2025cli}. By determining the probability of producing two or three particles in the same event with that from mixed events (where the particles are not correlated), the correlation function of these particles can be built. Once determined, information related to the scattering of these particles can be obtained, such as the scattering length and the effective range of the interaction~\cite{ALICE:2023bny,Liu:2024uxn}.  At the same time, calculation of invariant mass distributions in weak decays, determination of partial decay widths or investigating observables like cross sections can also reveal the exotic nature of states produced during the scattering~\cite{Brandao:2025cli,Lyu:2026rsm,Jia:2026dpl,Malabarba:2020grf,Malabarba:2023zez,Ren:2019umd,Ren:2019rts}.

 Here, we are going to focus on the properties of axial mesons with charm and strangeness. In particular, we study the properties of $D_1(2420)$, $D_{s1}(2460)$ and $D_{s1}(2536)$ as states generated from two-hadron interactions~\cite{Kolomeitsev:2003ac,Hofmann:2003je,Gamermann:2007fi,Cleven:2010aw,Malabarba:2022pdo}. In the case of $D_1(2420)$, the correlation function of $D^*\pi$ is investigated as well as the corresponding invariant mass distribution obtained from $B$ decays~\cite{Abreu:2024qqo,Brandao:2025cli}. For the case of the $D_{s1}(2460)$ and $D_{s1}(2536)$ states, their interaction with a nucleon is first investigated, finding the formation of three-body states, for later obtaining the corresponding $n\bar D_{s1}$ correlation functions~\cite{Agatao:2025ckp}.


\section{Correlation function and Invariant mass distributions related to $D_1(2420)$}
The properties of $D_1(2420)$ can be understood as a consequence of the interaction between pseudoscalars and vector mesons in the $s$-wave~\cite{Kolomeitsev:2003ac,Hofmann:2003je,Gamermann:2007fi,Cleven:2010aw,Malabarba:2022pdo}. Here, we follow the approach of Refs.~\cite{Gamermann:2007fi,Malabarba:2022pdo}, where channels like $D^*\pi$, $D\rho$ and $D_s \bar K^*$ are considered when solving the Bethe-Salpeter equation. In addition to this, in Ref.~\cite{Malabarba:2022pdo}, the decays $D^*\to D\pi$ and $\rho\to\pi\pi$ are implemented, producing a box diagram that contributes mainly to the amplitude $D\rho\to D\rho$. In this way, a pole $\sim 2428$ MeV with a total width of $\sim$ 33 MeV is found with spin-parity $J^P=1^+$ in the $T$-matrix determined from the resolution of the Bethe-Salpeter equation. This pole, which strongly couples to $D\rho$, can be related to $D_1(2420)$. Along the pole mentioned, another one, broader (width $\sim 122$ MeV) and with a mass of $\sim 2222$ MeV is obtained in Ref.~\cite{Malabarba:2022pdo}. The latter state strongly couples to $D^*\pi$, which is open for decay, a property that is in common with that observed for $D_1(2430)$. However, the fact that the mass found is too low to be related to that of $D_1(2430)$ suggests that this state could have a strong admixture of $q\bar q$ and a molecular $D^*\pi$ components. In Ref.~\cite{Malabarba:2022pdo} the latter possibility was investigated in relation to the discrepancy obtained for the $D\pi$, $D^*\pi$ scattering lengths between lattice QCD studies, chiral theories and lattice QCD-inspired models, where Bethe-Salpeter equations are solved using a kernel based on the chiral theory for heavy mesons and the parameters are fixed by fitting the scattering length obtained from lattice QCD calculations~\cite{Mohler:2012na,Guo:2018tjx,Guo:2009ct,Du:2017zvv}. Also, recently, the ALICE collaboration has used correlation functions to determine the $D^*\pi$ scattering length~\cite{ALICE:2024bhk}. 
\subsection{Correlation function}
In the case of a pair of particles, the correlation function 
$C(k)$, where $k$ is the modulus of the relative momentum between the particles in the center-of-mass-frame, can be written as~\cite{Koonin:1977fh,Pratt:1986cc}
\begin{align}
C(k)=\int d^3 r S_{12}(\vec{r})|\psi(\vec{k},\vec{r})|^2 \label{Ck}   
\end{align}
In Eq.~(\ref{Ck}), $\psi(\vec{k},\vec{r})$ is the wave function describing the relative motion between the particles, and $S_{12}$ represents the hadron source, related to the production of particles from the high-energy collision. Typically, a normalized static Gaussian source function depending on the source size parameter, $R$, is considered:
\begin{align}
S_{12}(\vec{r})=\frac{1}{(4\pi)^{3/2}R^3}e^{-r^2/(4R^2)}.
\end{align}
The wave function in Eq.~(\ref{Ck}) can be related to the coupled-channel $T$-matrix of the system under study, such that for a channel $i$~\cite{Vidana:2023olz},
\begin{align}
C_i(k)=1+4\pi\Theta(q_\text{max}-k)\int\limits_0^\infty dr r^2 S_{12}(\vec{r})\Bigg(\sum\limits_j w_j|j_0(kr)\delta_{ji}+T_{ji}(\sqrt{s})G_j(r,s)|^2-j^2_0(kr)\Bigg).\label{Ck2}   
\end{align}
In Eq.~(\ref{Ck2}), $\Theta$ is the Heaviside step function, $j_0$ is the spherical Bessel function $j_l$ for $l=0$, $w_j$ is the weight for channel $j$ (here we chose $w_j=1$), $T_{ji}$ is the two-body $T$-matrix describing the transition $j\to i$ and $G_j$ is given (for the case of two mesons) by
\begin{align}
G_j(r;s)=\int\limits_0^{q_\text{max}}\frac{d^3 q}{(2\pi)^3}\frac{\omega^{(j)}_1+\omega^{(j)}_2}{2\omega^{(j)}_1 \omega^{(j)}_2}\frac{j_0(qr)}{s-(\omega^{(j)}_1+\omega^{(j)}_2)^2+i\epsilon}.\label{G}
\end{align}
In Eq.~(\ref{G}), $\omega^{(j)}_a=\sqrt{k^2+m^2_a}$ represents the energy  of particle $a$ and $q_\text{max}$ is a momentum cut-off ($\sim 1000$ MeV) needed to regularize the integral when $r\to 0$.

\begin{figure}
\centering
\includegraphics[width=\textwidth]{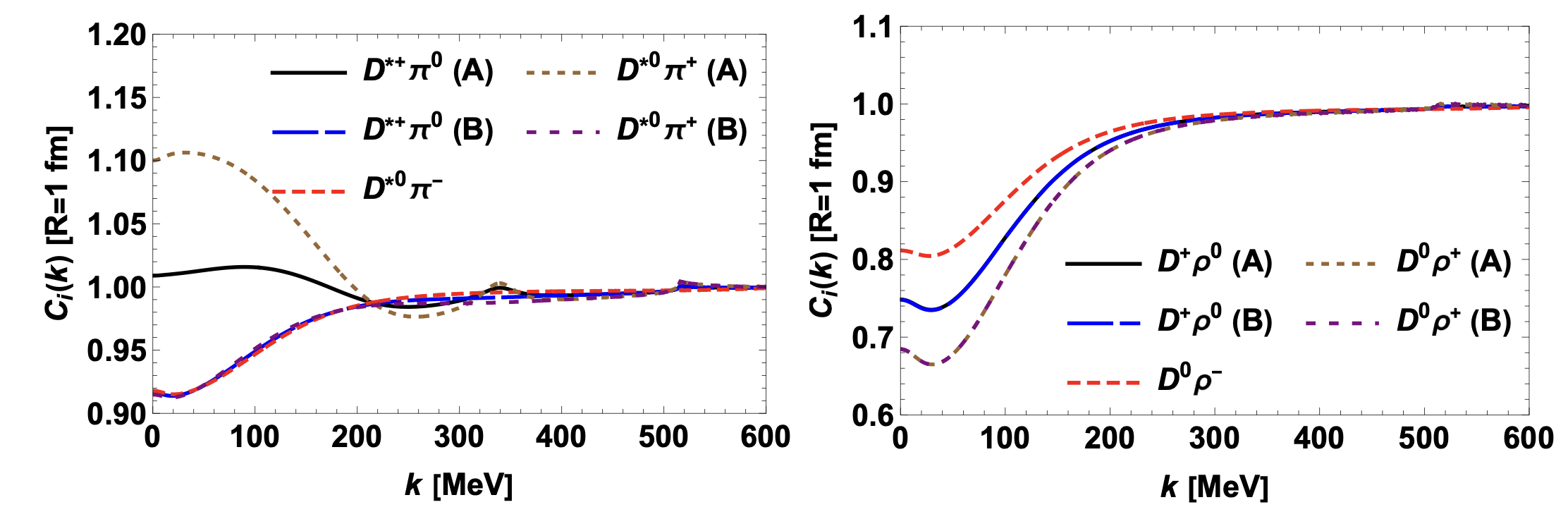}
\caption{Correlation functions for $D^*\pi$ and $D\rho$ in the charge basis. The source size is chosen in this case to be $R=1$ fm.}\label{cor}
\end{figure}

Next, we present our results for the $D^{*+(0)}\pi^{0(+)}$ correlation functions, where Coulomb interactions are absent. By introducing a pole related to a bare $q\bar q$ component in the $D^*\pi$ amplitude, the coupling constant and mass of this pole are varied to try to reproduce the mass and width of $D_1(2430)$ while finding a scattering length compatible with either that from lattice QCD-inspired models or from the ALICE collaboration. The impact of these solutions on the $D^{*+(0)}\pi^{0(+)}$ correlation functions is then studied.
Figure~\ref{cor} shows the correlation functions $C_{D^{*0}\pi^+}\equiv C_{D^{*0}\pi^+\to D^{*0}\pi^+}+C_{D^{*+}\pi^0\to D^{*0}\pi^+}$, $C_{D^{*+}\pi^0}\equiv C_{D^{*0}\pi^+\to D^{*+}\pi^0}+C_{D^{*+}\pi^0\to D^{*+}\pi^0}$, $C_{D^{0}\rho^+}$, $C_{D^{+}\rho^0}$ (with the latter defined analogously to the corresponding correlation functions for $D^*\pi$).
As can be seen in the figure, the correlation function distinguishes between the results related to a situation where the scattering length obtained for $D^*\pi$ is compatible with that of lattice QCD inspired models (case A) or with that obtained by the ALICE collaboration (case B).

\subsection{Invariant mass distribution}
\begin{figure}[h!]
    \centering
    \includegraphics[width=0.7\textwidth]{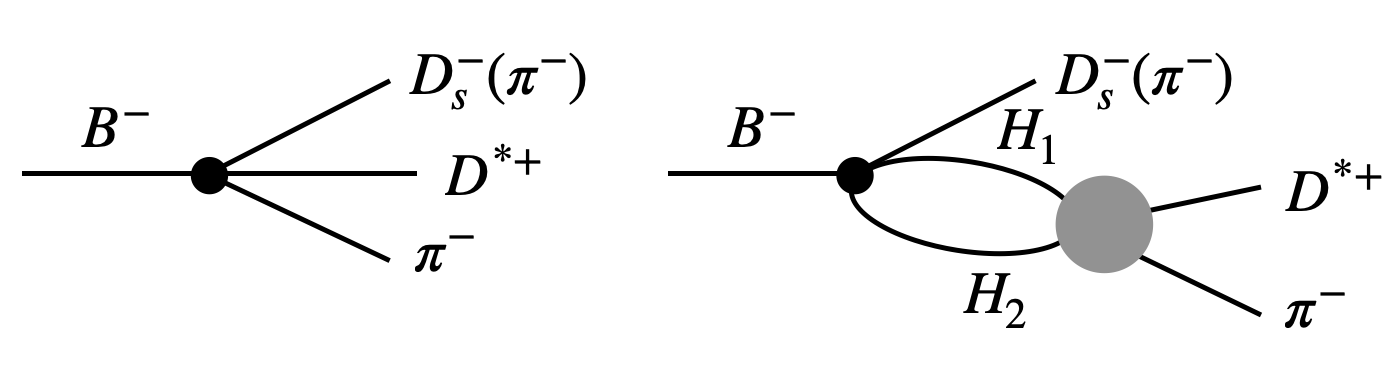}
    \caption{Decay mechanisms for the weak processes investigated.}
    \label{treeR}
\end{figure}

Recently, the LHCB collaboration has determined the invariant mass distribution of $D^{*+}\pi^-$ considering two weak processes: (the charged conjugation of) $B^-\to D^-_s D^{*+}\pi^-$ and $B^-\to D^{*+}\pi^-\pi^-$~\cite{LHCb:2024vhs,LHCb:2019juy}. In both cases, a clear signal for $D_1(2420)$ is observed. Considering that $D_1(2420)$ is generated from the $s$-wave dynamics of $D^*\pi$, $D\rho$ and other coupled channels~\cite{Malabarba:2022pdo}, the final states observed in these $B$ decays can be produced (see Fig.~\ref{treeR}) from a tree-level amplitude $B^-\to D^-_s H_1 H_2$ or $B^-\to H_1 H_2 \pi^-$, where $H_1 H_2$ denotes a pair of hadrons coupled to $D_1(2420)$, followed by the rescattering $H_1 H_2\to D_1(2420)\to D^*\pi$ (for the tree-level amplitude, $H_1 H_2=D^{*+}\pi^-$). 

The hadrons $H_1 H_2$ are obtained from a hadronization of the $q\bar q$ pairs involved in the weak transitions, considering external and internal emissions of a $W^-$. Once the amplitude considering the diagrams of Fig.~\ref{treeR} is calculated, the invariant mass distribution $d\Gamma/dm_{D^*\pi}$ (here $\Gamma$ is the decay width for the process studied and $m_{D^{*+}\pi^-}$ represents the $D^{*+}\pi^-$ invariant mass) can be obtained (in the case of having two $\pi^-$ in the final state, a symmetrized amplitude is used). For more details on the formalism, we refer the reader to Ref.~\cite{Brandao:2025cli}. 

\begin{figure}[h!]
    \centering
    \includegraphics[width=0.7\textwidth]{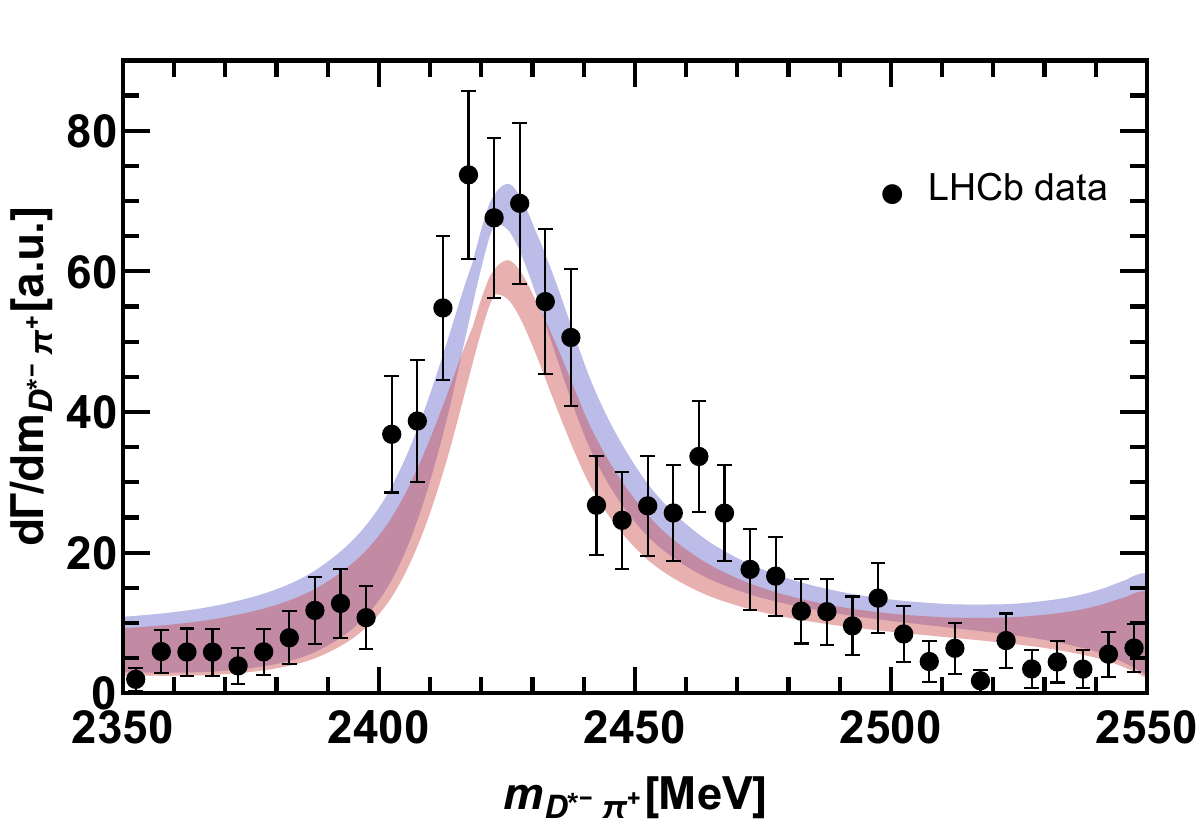}
    \caption{The $D^{*+}\pi^-$ invariant mass distribution. The data are taken from Ref.~\cite{LHCb:2024vhs}.}
    \label{Dstarpi1}
\end{figure}

In Fig.~\ref{Dstarpi1}, we show the results found for the $D^{*+}\pi^-$ invariant mass distribution of the process $B^-\to D^-_s D^{*+}\pi^-$. As can be seen, a model where $D_1(2420)$ is generated from pseudoscalar-vector dynamics in the $s$-wave can explain quite well the behavior observed by the LHC collaboration. 
As can be seen, there are two bands shown in Fig.~\ref{Dstarpi1} together with the experimental data. These bands are obtained by varying the cut-off and form factor used to regularize the loop integrals appearing in the evaluation of the rescattering contributions shown in Fig.~\ref{treeR}. The band associated with a bigger (smaller) value of $d\Gamma/dm_{D^{*+}\pi^-}$  is obtained by fixing an arbitrary constant present in the amplitude of the diagrams in Fig.~\ref{Dstarpi1} to reproduce $100$ ($85$) $\%$ of the total area (considering the central values) of the experimental distribution. The mentioned constant is a consequence of the fact that the hadronization procedure considered fixes only the relative strength between different final states. We refer the reader to Ref.~\cite{Brandao:2025cli} for the results obtained for $B^-\to D^{*+}\pi^-\pi^-$ and for more discussion on other invariant mass distributions, such as $D^-_s D^{*+}$, $\pi^-\pi^-$, of the investigated reactions.

\section{Formation of three-body states in the $n\bar D_{s1}$ system}
In Ref.~\cite{Gamermann:2007fi}, the properties of the states $D_{s1}(2460)$, $D_{s1}(2536)$ could be understood in terms of pseudoscalar-vector meson dynamics, such as $D^*K$, $DK^*$ and coupled channels, in the $s$-wave. At the same time, the $\bar K N$ and $\bar K^* N$ interactions are attractive in the $s$-wave as well, generating $\Lambda(1405)$ and $\Lambda(1800)$, respectively,~\cite{Kaiser:1995eg,Oset:1997it,Garzon:2012np}. In view of these attractive interactions, considering a system like $N\bar D^*\bar K$ or $N\bar D\bar K^*$ could lead to the formation of three-body states.

\begin{figure}
    \centering
    \includegraphics[width=0.3\textwidth]{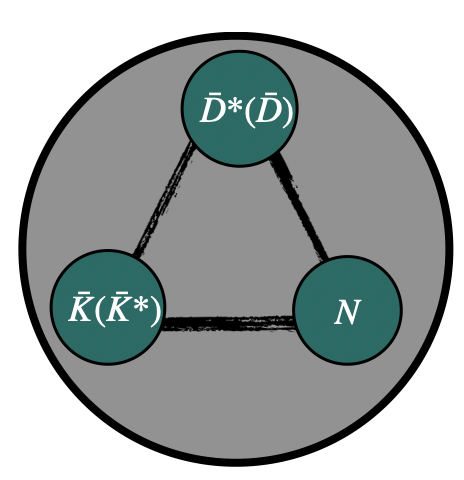}
    \caption{Three-body states originated from the $s$-wave interactions in the  $N\bar D^*\bar K$ or $N\bar D\bar K^*$ systems.}
    \label{fig:placeholder}
\end{figure}
To investigate such a possibility, we determine the $T$-matrix for these systems considering that $\bar D^* \bar K$ ($\bar D \bar K^*$) clusters as $\bar D_{s1}(2460)$ [$\bar D_{s1}(2536)$]. To do this, we use the so-called fixed center approximation to the Faddeev equations, where the interaction of the three-particles resembles to that of a particle with a fixed scattering center~\cite{MartinezTorres:2020hus}. In this way, considering first the successive rescattering of $N$ with the particles that constitute the cluster, the amplitudes $T_{ij}$, $i,j=1,2$, are obtained. In this context, $T_{ij}$ represents contributions where $N$ interact first with particle $i$ of the cluster and, after successive rescattering with both constituents of the cluster, ends up interacting with particle $j$ of the cluster. Mathematically, these amplitudes can be obtained by solving the coupled equations~\cite{Agatao:2025ckp}:

\begin{align}
T_{ij}=t_i\delta_{ij}+t_i g T_{kj}\label{fca}
\end{align}
with $i,j,k=1,2$ and $k\neq i$. In Eq.~(\ref{fca}), $t_i$ represents the two-body $t$-matrix of $N$ with particle $i$ of the cluster, and $g$ corresponds to the propagator of $N$ in the cluster. This $g$ contains a form factor which depends on the molecular nature of the cluster. We refer the reader to Ref.~\cite{Agatao:2025ckp} for more details. The amplitudes $T_{ij}$ are then used as a kernel in a Lippmann-Schwinger type equation to implement the particle-cluster propagation in each of the contributions considered in Eq.~(\ref{fca}):
\begin{align}
    \mathcal{T}=T+T \mathcal{G}\mathcal{T},\label{LS}
\end{align}
where 
\begin{align}
    T=\left(\begin{array} {cc}T_{11}&T_{12}\\T_{21}&T_{22}\end{array}\right),~\mathcal{G}=\left(\begin{array} {cc}\mathcal{G}^{(1)}&0\\0&\mathcal{G}^{(2)}\end{array}\right).\label{TG}
\end{align}
In Eq.~(\ref{TG}), $\mathcal{G}^{(i)}$ represents the particle-cluster propagator. This propagator, as in the case of $g$ in Eq.~(\ref{fca}), depends on a form-factor related to the molecular nature of the cluster. The superscript $(i)$ in this propagator refers to the particle $i$ of the cluster, indicating a different momentum configuration in this form factor as a consequence of the collision of the nucleon with particle $i$ of the cluster. The $T$-matrix describing the particle-cluster interaction is then given by
\begin{align}
\mathcal{T}=\sum\limits_{i,j=1}^2 \mathcal{T}_{ij},    
\end{align}
and it satisfies elastic two-body unitarity.
\begin{figure}
\centering
\includegraphics[width=0.6\textwidth]{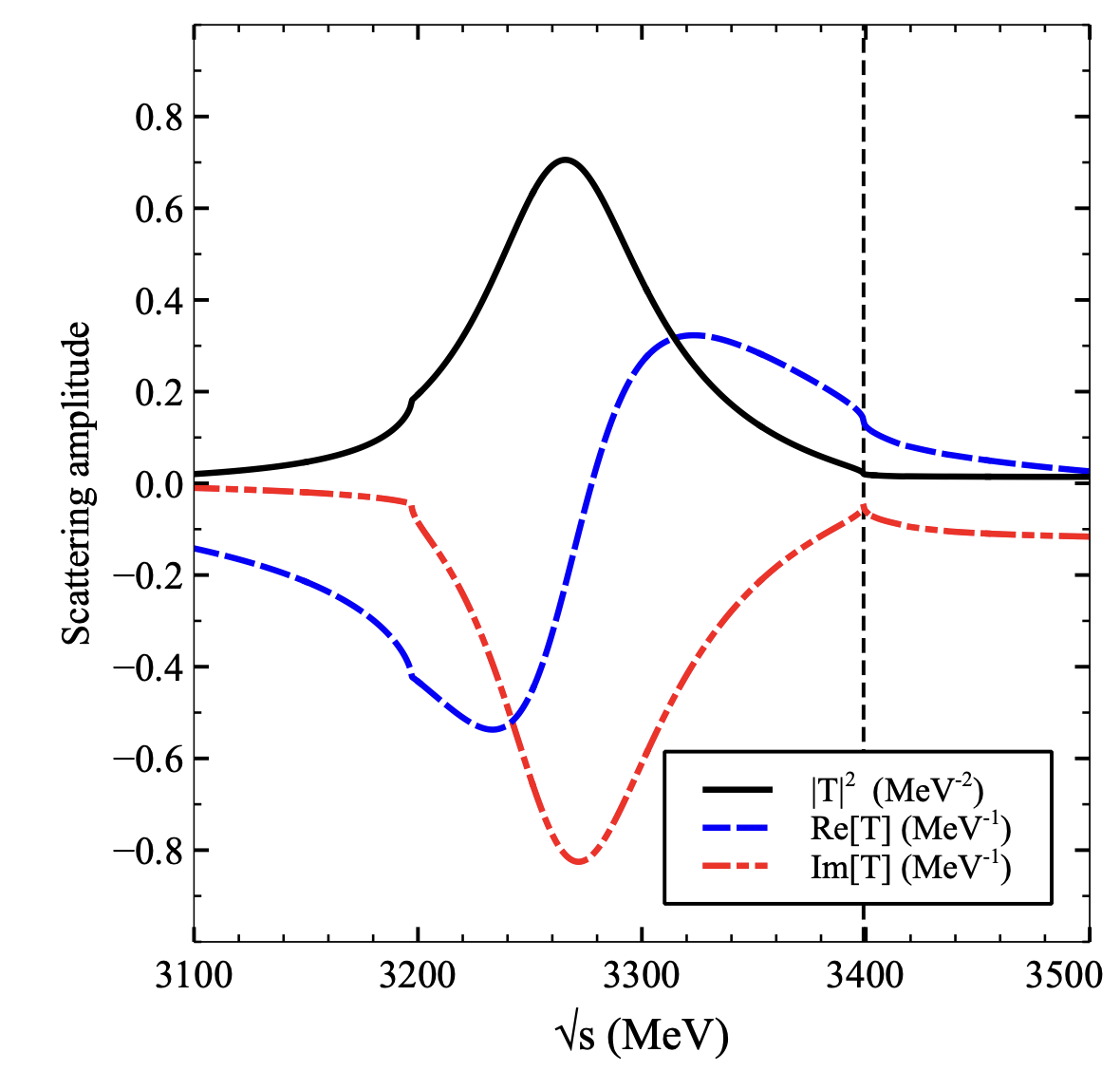}
\caption{$T$-matrix describing $n\bar D_{s1}(2460)\to n\bar D_{s1}(2460)$. The vertical line represents the particle+cluster threshold.}\label{T3}
\end{figure}

In Fig.~\ref{T3}, we show the results obtained for the $T$-matrix of the $n\bar D_{s1}(2460)$ system, where $n$ represents the neutron. We choose a neutron to avoid Coulomb interactions when calculating the correlation function of the system. As can be seen, a state with a mass of $\sim 3265$ MeV is obtained with a width of $\sim$ 90 MeV. The origin of this width is in the input two-body $t$-matrices appearing in Eq.~(\ref{fca}), which are obtained within a coupled channel approach and an imaginary part is produced.

Using the $T$-matrix obtained, the correlation function for the system can be determined from Eq.~(\ref{Ck}) considering only one channel and changing $T_{ji}G_j$ by $(\mathcal{T}_{11}+\mathcal{T}_{12})\tilde{G}_1+(\mathcal{T}_{12}+\mathcal{T}_{22}\tilde{G}_2)$~\cite{Agatao:2025ckp}. Here, $\tilde{G}_i$, $i=1,2$, is analogous to $G_j$ in Eq.~(\ref{Ck}) but it includes a form-factor describing the cluster and a $2m_n$ normalization factor related to the neutron. The form factor mentioned is the same as the one in $\mathcal{G}^{(i)}$ of Eq.~(\ref{LS}).
\begin{figure}[h!]
\centering
\includegraphics[width=0.5\textwidth]{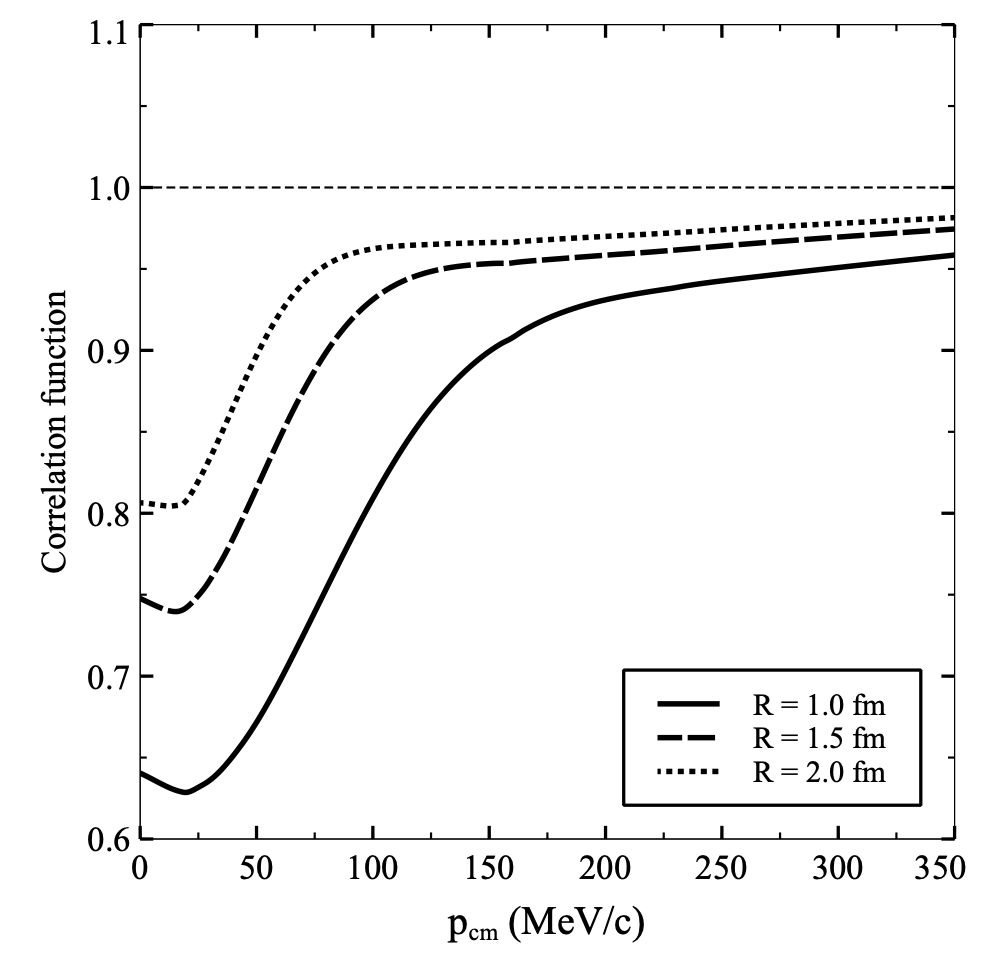}
\caption{Correlation function of the $n\bar D_{s1}(2460)$ system.}\label{C3}
\end{figure}

In Fig.~\ref{C3}, we show the correlation function for the mentioned system as a function of the center-of-mass momentum. A depletion at low momentum and a growth as we go further in momentum is observed, getting closer to unity. We also observe that an increase in $R$ (the size of the source) produces a larger magnitude for the correlation function at low momentum. The behavior observed is typical for states below the threshold of the system, as it is the case here.

Similar results are found for the $n\bar D_{s1}(2536)$ system.
\section*{Acknowledgements}
This work is partly supported by the Spanish Ministerio de Economia y Competitividad (MINECO) and European FEDER funds under Contracts No. FIS2017- 84038-C2-1-P B, PID2020- 112777GB-I00, and by Generalitat Valenciana under contracts PROMETEO/2020/023 and CIPROM/2023/59. This project has received funding from the European Union Horizon 2020 research and innovation programme under the program H2020- INFRAIA2018-1, grant agreement No. 824093 of the STRONG-2020 project. This work is supported by the Spanish Ministerio de Ciencia e Innovaci\'on (MICINN) under contracts PID2020-112777GB-I00, PID2023-147458NBC21 and CEX2023-001292-S. A.M.T,
K.P.K, L.M.A, B.A and P.B gratefully acknowledge the partial support provided by the Brazilian agency CNPq (K.P.K: Grants No. 407437/ 2023-1 and No. 306461/2023-4; A.M.T: Grant No. 304510/2023-8, L.M.A.: Grants No. 400215/2022-5, 308299/2023-0, 402942/2024-8, B.A: Grant No. 987654/2023-1 and 200204/2025-4, P.B: Grant No. 200203/2025-8). Further, L. M. A partly acknowledges the support from CNPq/FAPERJ under the Project INCT-F\'isica Nuclear e Aplica\c c\~oes (Contract No. 464898/2014-5).

\end{document}